\documentclass[%
  manuscript=letter,
  journal=jpclcd
]{achemso}

\usepackage[utf8]{inputenc}
\usepackage[T1]{fontenc}
\usepackage{xcolor}
\definecolor{blue}{RGB}{0, 0, 139}
\usepackage[%
  colorlinks=true,
  allcolors=blue
]{hyperref}
\usepackage{url}
\usepackage{enumitem}
\usepackage{amsfonts}
\usepackage{amssymb}
\usepackage{amsmath}
\usepackage{mathtools}
\usepackage{lmodern}

\DeclarePairedDelimiter\ppar{(}{)}              % ( )
\DeclarePairedDelimiter\pnrm{\lVert}{\rVert}    % || ||
\DeclarePairedDelimiter\pset{\{}{\}}            % { }

\newcommand{\rfig}[1]{Figure~\ref{#1}}
\newcommand{\rref}[1]{ref~\citenum{#1}}
\newcommand{\req}[1]{eq~\ref{#1}}

\newcommand{\dd}[1]{\operatorname{d#1}}
\newcommand{\bx}{\mathbf{x}}
\newcommand{\bz}{\mathbf{z}}
\newcommand{\dx}{\dd{\mathbf{x}}}
\newcommand{\e}{\operatorname{e}}
\newcommand{\btheta}{\boldsymbol{\theta}}

\newcommand{\note}[1]{{\color{black}{#1}}}

\title{Spectral Map: \\ Embedding Slow Kinetics in Collective Variables}

\author{Jakub Rydzewski}
\email{jr@fizyka.umk.pl}
\affiliation{%
  Institute of Physics,
  Faculty of Physics, Astronomy and Informatics,
  Nicolaus Copernicus University,
  Grudziadzka 5, 87-100 Toru\'n, Poland
}

\begin{document}

\begin{tocentry}
  \begin{center}
    \includegraphics{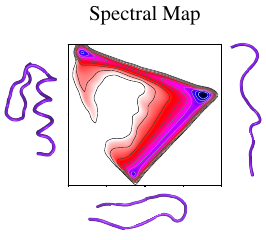}
  \end{center}
\end{tocentry}

\begin{abstract}
The dynamics of physical systems that require high-dimensional representation can often be captured in a few meaningful degrees of freedom called collective variables (CVs). However, identifying CVs is challenging and constitutes a fundamental problem in physical chemistry. This problem is even more pronounced when CVs information about slow kinetics related to rare transitions between long-lived metastable states. To address this issue, we propose an unsupervised deep-learning method called spectral map. Our method constructs slow CVs by maximizing the spectral gap between slow and fast eigenvalues of a transition matrix estimated by an anisotropic diffusion kernel. We demonstrate our method in several high-dimensional reversible folding processes.
\end{abstract}

\maketitle

\note{
In physical chemistry, identifying slowly varying order parameters known as collective variables (CVs) in complex systems is essential~\cite{rydzewski2023manifold,rohrdanz2013discovering,valsson2016enhancing,ceriotti2019unsupervised}. However, this challenging task often requires relying on physical intuition or trial and error. For processes exhibiting many timescales, such as glass transitions~\cite{van2021towards} or crystallization~\cite{neha2022collective}, CVs should capture the slow dynamics of rare transitions between long-lived metastable states. Failure to encode this can hinder the description of the underlying physical mechanisms. To address this, several techniques for learning CVs directly from simulation data have been developed~\cite{singer2009detecting,ceriotti2011simplifying,noe2015kinetic,chiavazzo2017intrinsic,zhang2018unfolding,rydzewski2020multiscale,bonati2020data,raucci2022discover,rydzewski2022reweighted,evans2023computing,jung2023machine}.

In many metastable systems, the long-term dynamics can often be approximately Markovian and modeled as diffusion along CVs in the presence of a free-energy landscape~\cite{berezhkovskii2011time}. The timescale separation between informative slow and hidden fast processes can be ensured by maximizing the spectral gap between the slow and fast eigenvalues of the Markov transition matrix~\cite{berezhkovskii2011time,coifman2008diffusion,tiwary2016spectral,pmlr-v108-pillaud-vivien20a}. This results in the adiabatic elimination of fast variables, which are slaved to the slow variables or their statistics, leading to the effective slow dynamics of the system. However, selecting inappropriate CVs can result in non-Markovian dynamics with long memory effects~\cite{zwanzig1961memory}.

In this Letter, we propose spectral map, a straightforward method that can construct CVs that arise due to the adiabatic timescale separation in complex systems. Spectral map draws inspiration from recent developments in diffusion maps and parametric dimensionality reduction techniques~\cite{rydzewski2022reweighted,rydzewski2023selecting,rydzewski2023manifold}. Our method is based on training a neural network to embed a high-dimensional system into a few slow CVs without supervision. Maximizing the spectral gap, we parametrize a low-dimensional representation corresponding to the slowest timescales relevant to the physical system.}

In the following, we consider a high-dimensional system described by $n$ configuration variables $\bx = (x_1, \dots, x_n)$ whose dynamics at temperature $T$ is driven according to a potential energy function $U(\bx)$, and sampled generally from an unknown equilibrium distribution. However, if we represent the system using the microscopic coordinates, its dynamics proceeds according to a canonical equilibrium distribution given by the Boltzmann density $p(\bx)=\e^{-\beta U(\bx)}/Z$, where $\beta=(k_{\mathrm{B}}T)^{-1}$ is the inverse temperature and $Z=\int\dx\e^{-\beta U(\bx)}$ is the partition function of the system.

We simplify the high-dimensional configuration space by mapping it into a reduced space $\bz=\ppar*{z_1, \dots, z_d}$ given by a set of $d$ functions of the configuration variables, commonly referred to as CVs, where $d \ll n$. Furthermore, we can encapsulate these functions into a parametrizable target mapping~\cite{rydzewski2020multiscale,rydzewski2022reweighted,rydzewski2023manifold}:
\begin{equation}
  \label{eq:target-mapping}
  \bz = \xi_{\theta}(\bx) \equiv \pset[\big]{\xi_k(\bx, \theta)}_{k=1}^d,
\end{equation}
where $\theta$ are adjustable parameters. The target mapping ensures that CVs represent the slowest variables in the system. By sampling the system in the CV space, its dynamics follows a marginal equilibrium density $p(\bz)\propto\e^{-\beta F(\bz)}$, where $F(\bz)$ is a free-energy landscape:
\begin{equation}
  F(\bz) = -\frac{1}{\beta}\log\int\dx\,\delta\ppar*{\bz - \xi_\theta(\bx)}\e^{-\beta U(\bx)},
\end{equation}
up to an irrelevant constant, and $\delta(\cdot)$ is the Dirac delta function.

To estimate the effective timescales characteristic of the system, we can model its dynamics as a Markov chain using kernel functions. However, as the marginal equilibrium density $p(\bz)$ is unknown, we need a density-preserving kernel appropriate for data sampled from any underlying probability distribution. To achieve this, we employ an anisotropic diffusion kernel~\cite{coifman2005geometric,coifman2006diffusion,coifman2008diffusion}:
\begin{equation}
  \label{eq:kappa}
  \kappa(\bz_k,\bz_l) = \frac{g(\bz_k,\bz_l)}{\sqrt{\varrho(\bz_k)\varrho(\bz_l)}},
\end{equation}
where $g(\bz_k,\bz_l)=\exp\ppar*{-\pnrm{\bz_k-\bz_l}^2/\varepsilon}$ is a Gaussian kernel used to describe pairwise Euclidean distances $\|\bz_k-\bz_l\|$ between CV samples, $\varrho(\bz_k)=\sum_l g(\bz_k,\bz_l)$ is a kernel density estimate, and $\varepsilon$ is a scale constant. Then, a Markov transition matrix can be constructed by normalizing \req{eq:kappa}:
\begin{equation}
  \label{eq:mmm}
  m_{kl} \sim M(\bz_k,\bz_l) = \frac{\kappa(\bz_k, \bz_l)}{\sum_n \kappa(\bz_k, \bz_n)},
\end{equation}
which describes a Markov chain in the CV space given by $m_{kl}=\mathrm{Pr}\,\pset{\bz_{t+1}=\bz_l\,|\, \bz_t = \bz_k}$ that corresponds to a transition probability from $\bz_k$ to $\bz_l$ in an auxiliary time $t$. 

\note{
As the Markov transition matrix is estimated in the reduced space, we consider the effective dynamics rather than the dynamics of the microscopic coordinates of the system. Thus, the Markov chain defined in the CV space is implicitly modeled as following an overdamped Langevin dynamics with the marginal equilibrium density $p(\bz)$. Generally, this effective dynamics is either non-Markovian or has a $\bz$-dependent diffusion matrix, which means that it is not driven exclusively by the free-energy landscape~\cite{maragliano2006string,legoll2010effective,zhang2016effective}. However, by selecting CVs that arise from the timescale separation in the system as the reduced space, we can represent the dynamics of microscopic coordinates through the effective dynamics of slow CVs, which is approximately Markovian~\cite{berezhkovskii2011time,tiwary2016spectral}.

Consequently, to ensure that the timescale separation is evident and that CVs closely follow the slow dynamics of the system, we perform a spectral decomposition of the Markov transition matrix in the CV space.} This allows us to calculate its dominant eigenvalues $\lambda_0 = 1 \ge \lambda_1 \ge \lambda_2 \ge \dots$. The difference between neighboring eigenvalues is called the spectral gap and measures the degree of the timescale separation between the slow and fast variables:
\begin{equation}
  \label{eq:spectral-gap}
  \sigma = \lambda_{k-1} - \lambda_k
\end{equation}
where $k > 0$ indicates the number of metastable states in the CV space. As such, our objective is to reach the largest spectral gap between the slow and fast variables by adjusting the parameters of the target mapping $\xi_\theta$ (\req{eq:target-mapping}).

\begin{figure}[t]
  \includegraphics{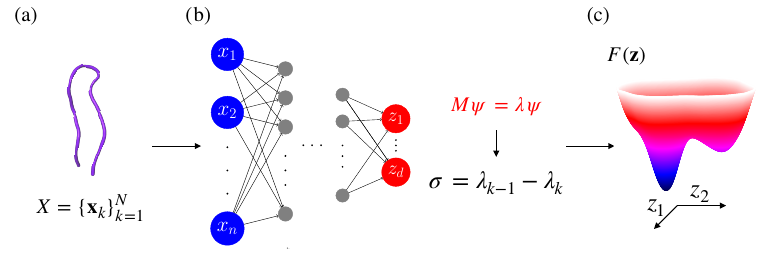}
  \caption{Outline of spectral map. (a) Dataset $X$ in a high-dimensional representation $\bx=\ppar{x_1,\dots,x_n}$ used to describe the system is taken as an input for the target mapping. (b) Target mapping $\bz=\xi_{\btheta}(\bx)$ is modeled as a neural network that embeds the system in its high-dimensional representation to a low-dimensional map spanned by slow CVs $\bz=\ppar{z_1,\dots,z_d}$. \note{An eigendecomposition of a Markov transition constructed from CV samples is performed ($M\psi=\lambda\psi$).} The spectral gap $\sigma$ is maximized based on the difference between neighboring eigenvalues $\pset{\lambda_k}$ to separate the slow and fast timescales. (c) Trained neural network can be used to evaluate all available high-dimensional samples and calculate the corresponding free-energy landscape $F(\bz)$.}
  \label{fig:spectral-map}
\end{figure}

Let us summarize the algorithm to calculate spectral map:
\begin{enumerate}
  \item A dataset consisting of $N$ configuration samples in a high-dimensional representation, $X=\pset*{\bx_k}_{k=1}^{N}$, is obtained from a simulation; see \rfig{fig:spectral-map}(a).
  
  \item The target mapping $\xi_\theta(\bx)$ represented as a neural network is trained through backpropagation by feeding the dataset $X$ and mapping it to $d$ CVs. The objective is to maximize the spectral gap estimated from the eigendecomposition of the Markov transition matrix constructed from CV samples; see \rfig{fig:spectral-map}(b).
  
  \item The trained target mapping is used to evaluate $N$ samples $\pset*{\bz_k}_{k=1}^{N}$ in the CV space and estimate the corresponding free-energy landscape $F(\bz)$; see \rfig{fig:spectral-map}(c).
\end{enumerate}
\begin{figure}[t]
  \includegraphics{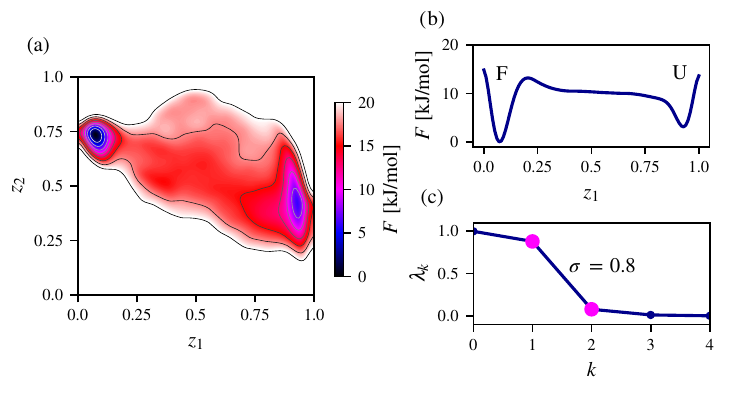}
  \caption{Spectral map of chignolin folding in solvent calculated from a training dataset consisting of 5000 samples extracted every 2 ns from a 100-$\mu$s molecular dynamics simulation at a temperature of 340 K. The high-dimensional representation of chignolin is given by $n=45$ pairwise Euclidean distances between C$\alpha$ atoms. (a) Free-energy landscape estimated from the complete trajectory (sampled every 200 ps). The folded state is the main metastable state, with a less populated state representing the unfolded state. The states are separated by a free-energy barrier of around 12 kJ/mol. (b) Free-energy profile along the $z_1$ CV with $z_2$ integrated out. (c) The maximal separation between eigenvalues is obtained for the spectral gap of 0.8 at $k=2$ (\req{eq:spectral-gap}), with successive eigenvalues close to 0.}
  \label{fig:cln}
\end{figure}

As an initial application of our method, we consider the folding process of a ten-residue protein chignolin in solvent. A 100-$\mu$s unbiased molecular dynamics simulation at a temperature of 340 K of this process is obtained from \rref{lindorff2011fast}. As a high-dimensional representation for chignolin, we use pairwise Euclidean distances between its C$\alpha$ atoms, amounting to $n=45$ configuration variables. The training set consists of 5000 samples extracted from the simulation every 2 ns. The training of the target mapping is carried out using $k=2$ for 100 epochs with data batches consisting of 100 high-dimensional samples. Once the target mapping is trained, we evaluate all samples collected from the simulation (sampled every 200 ps) to construct the corresponding free-energy landscape. For additional technical details, we refer to the Supporting Information.

Our results are presented in \rfig{fig:cln}. By selecting $k=2$ for the spectral gap (\req{eq:spectral-gap}), we identify two metastable states in the free-energy landscape spanned by CVs found by spectral map, as depicted in \rfig{fig:cln}(a). We can see that the deepest free-energy minimum corresponds to the folded ensemble of chignolin, while the less populated metastable state consists of an ensemble of unfolded conformations. This can also be observed along $z_1$ by integrating out $z_2$ in \rfig{fig:cln}(b). The folded and unfolded states are separated by a free-energy barrier of around 12 kJ/mol. From \rfig{fig:cln}(c), we can see that the spectral gap between the first and second eigenvalues of the corresponding Markov transition matrix is maximal at around $0.8$, with $\lambda_1 \sim 1$ as dominant and $\lambda_2$ and successive eigenvalues close to 0. This indicates that the metastable states are well-separated, and spectral map contains information about the slow kinetics of the folding process of chignolin.

As a following example, let us consider the reversible folding of trp-cage, which is simulated through a 200-$\mu$s molecular dynamics simulation at a temperature of 290 K in solvent (obtained from \rref{lindorff2011fast}). We use pairwise Euclidean distances between C$\alpha$ atoms ($n=190$ configuration variables) as a high-dimensional representation. The training set consists of 10000 samples extracted from the simulation every 2 ns. The training of the target mapping is carried out for $k=2$ to 7 over 100 epochs, with data batches consisting of 100 high-dimensional samples. Finally, the calculated spectral maps and free-energy landscapes are computed using the complete trajectory (sampled every 200 ps). For additional details, see the Supporting Information.

\begin{figure}[t]
  \includegraphics[width=\textwidth]{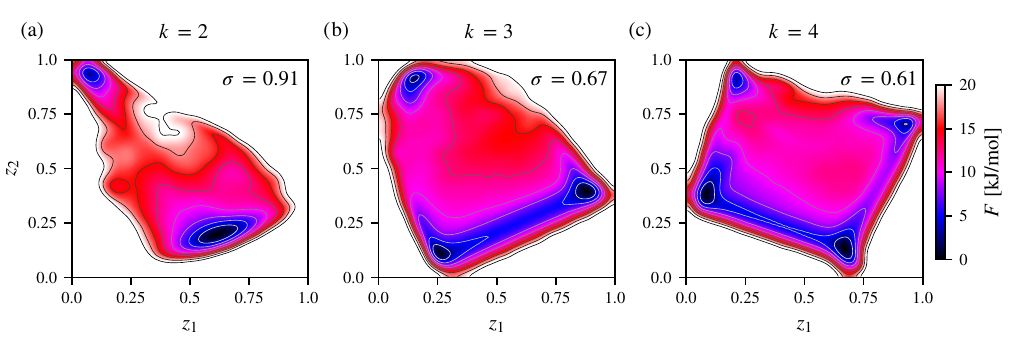}
  \caption{Spectral map and free-energy landscapes of the folding of trp-cage in solvent calculated from a training dataset consisting of 10000 samples extracted every 2 ns from a 200-$\mu$s molecular dynamics simulation at a temperature of 290 K. In total, $n=190$ pairwise Euclidean distances between C$\alpha$ atoms are used as a high-dimensional representation. The spectral gaps for $k=2, 3$, and 4 metastable states in (a), (b), and (c), respectively, indicate increasingly worsening separation between the timescales. In each spectral map, the unfolded state of trp-cage splits into additional metastable substates. The spectral gap values up to $k=7$ and a comparison of the CVs found by spectral map to CVs often used for reversible folding are shown in the Supporting Information.}
  \label{fig:trpcage}
\end{figure}

In this example, we aim to determine the optimal number of metastable states ($k$ in \req{eq:spectral-gap}) for maximizing the spectral gap. Our findings, presented in \rfig{fig:trpcage}, show spectral maps and corresponding free-energy landscapes for $k=2,3,4$ metastable states. As can be seen in \rfig{fig:trpcage}(a), selecting $k=2$ metastable states results in a distinct folded state and a loosely defined unfolded state. The spectral gap reaches its maximum value of $\sigma=0.91$, indicating satisfactory timescale separation. For $k=3$ and 4, the spectral gap decreases to $\sigma=0.67$ and 0.61, respectively, as shown in \rfig{fig:trpcage}(b-c). We observe that the folded state remains unchanged for $k=3$ and 4, while the unfolded state splits into several states due to its heterogeneity, resulting in worse timescale separation.

\note{Additionally, in \rfig{fig:trpcage}, we can observe that as the number of metastable states ($k$) increases and the timescale separation worsens, the difficulty in separating the metastable states also results in underestimated free-energy barriers. Based on this, we propose a criterium to determine the optimal value of $k$. Specifically, we can calculate several spectral maps for increasing values of $k$ and choose the one with the largest spectral gap, corresponding to the maximal separation between slow and fast eigenvalues. Moreover, such timescale separation induces the highest free-energy barriers between metastable states, indicating the superior quality of slow CVs. In the Supporting Information, we provide further details and demonstrate how the spectral gap decreases as the number of metastable states increases to $k=7$.}
\begin{figure}[t]
  \includegraphics{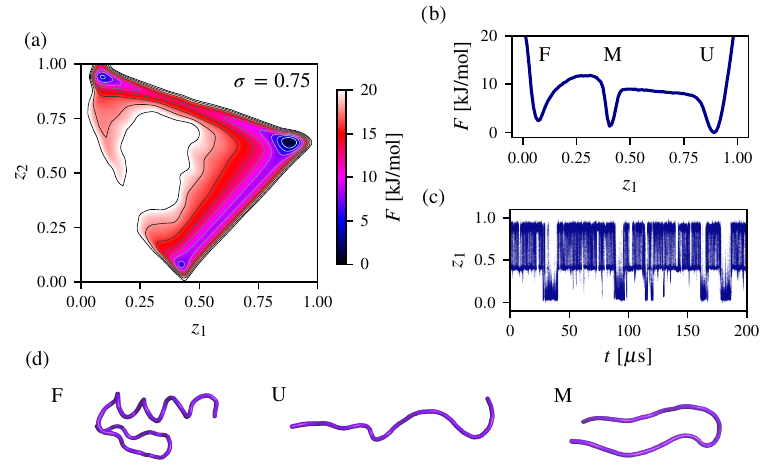}
  \caption{Spectral map and free-energy landscape of the folding of the BBA protein in solvent calculated from a training dataset consisting of 10000 samples extracted every 2 ns from a 200-$\mu$s molecular dynamics simulation at a temperature of 325 K. A high-dimensional representation given by $n=378$ pairwise Euclidean distances between the C$\alpha$ atoms of BBA. (a) Free-energy landscape showing metastable states spanned by CVs calculated by spectral map for $k=3$ where the corresponding spectral gap reaches $\sigma=0.75$. (b) Free-energy profile along the $z_1$ CV with $z_2$ integrated out. (c) Time series of $z_1$ showing changes between metastable states during the molecular dynamics simulation. (c) Representative conformations of the BBA protein corresponding to the folded state (F), the unfolded state (U), and the misfolded state in a $\beta$-hairpin structure (M).}
  \label{fig:bba}
\end{figure}

For our final example, we examine the folding and unfolding processes of the BBA protein in solvent, which is simulated by a 200-$\mu$s molecular dynamics simulation at a temperature of 325 K obtained from \rref{lindorff2011fast}. As previously, we consider pairwise Euclidean distances between the C$\alpha$ atoms of BBA ($n=378$ configuration variables) as its high-dimensional representation. The training set consists of 10000 samples extracted from the simulation every 2 ns. The training of the target mapping is carried out for $k=3$ metastable states through 100 epochs with data batches consisting of 100 high-dimensional samples. Finally, the spectral map and free-energy landscape are constructed using the complete trajectory (sampled every 200 ps). Additional details and protocol parameters are available in the Supporting Information.

Our results regarding the BBA protein are presented in \rfig{fig:bba}. We focus on the CVs and corresponding free-energy landscape for $k=3$ metastable states, as the maximal spectral gap for BBA is virtually the same for $k=2$. For the number of metastable states $k>3$, the separation between the slow and fast dynamics of BBA worsens (see the Supporting Information). The CVs found by spectral map uncover three distinct states, as shown in \rfig{fig:bba}(a-c). We find that the unfolded state, not the folded state, is the primary and most populated free-energy basin, suggesting that the folded state is relatively less stable in solvent. This has also been observed in \rref{lindorff2011fast}. Furthermore, we can see that spectral map also finds a $\beta$-hairpin state in addition to the folded and unfolded states of BBA. Representative conformations of BBA in these three states are shown in \rfig{fig:bba}(d).

\note{The capability of identifying the slow kinetics of the BBA protein by spectral map can also be demonstrated by comparing the learned slow CVs to physical descriptors used routinely for reversible folding processes, such as RMSD and the fraction of native contacts. As shown in the Supporting Information, these descriptors calculated in reference to the folded state of BBA are unable to distinguish between the unfolded and $\beta$-hairpin metastable states obtained by spectral map.}

\note{Interestingly, as can be seen in \rfig{fig:bba}(a), the minimum free-energy path found by spectral map corresponds to the transition directly between the folded and unfolded states. In contrast, the transition between the folded and $\beta$-hairpin states is less likely to be sampled for BBA. This cannot be observed by examining the free-energy profile shown in \rfig{fig:bba}(b), where the transition from the folded to unfolded states proceeds through the $\beta$-hairpin state. In contrast, the slow CVs found by spectral map reveal that the $\beta$-hairpin state is primarily reached from the unfolded state, indicating that this state is not an intermediate but a misfolded one.}

In this Letter, we have introduced spectral map. Our technique identifies slow CVs from high-dimensional observations obtained from a molecular dynamics simulation. Spectral map employs a procedure to separate slow and fast eigenvalues of a Markov transition matrix that is defined in the CV space. The timescale separation is obtained by maximizing the spectral gap by training a deep neural network. Spectral map allows us to encode information about slow kinetics in the physical system in just a few CVs.

\note{Spectral map shares conceptual similarities with several recent techniques, such as spectral gap optimization of order parameters (SGOOP)~\cite{tiwary2016spectral} and the variational approach for Markov processes in a deep learning framework (VAMPnet)~\cite{mardt2018vampnets}. However, there are notable differences. For example, SGOOP constructs a linear combination of trial CVs and a Markov chain from counting transition between states. VAMPnet employs eigenvalues derived from a spectral decomposition of a lag-time dependent correlation matrix for its loss function. In contrast, spectral map creates a Markov transition matrix from the anisotropic diffusion kernel, which does not require time series as input. This alternative approach could simplify the process of constructing slow CVs, eliminating the need to select a lag time in Markov state models that require time-dependent data to estimate the transition matrix.}

For demonstration purposes, we have shown a method for identifying slow CVs solely from unbiased molecular dynamics simulations. However, only a simple adjustment is needed to expand spectral map and learn from enhanced sampling simulations. This adjustment involves a reweighting procedure for unbiasing Markov transition matrices with statistical weights from a biased simulation. Such a procedure is implemented in reweighted diffusion maps~\cite{rydzewski2022reweighted,rydzewski2023manifold} and reweighted stochastic embedding techniques~\cite{zhang2018unfolding,rydzewski2020multiscale,rydzewski2022reweighted}. We intend to further explore this approach in subsequent works. 

Overall, spectral map shows promise in addressing the challenge of computing slow CV by effectively reducing the dimensionality of complex physical systems and has a large potential for further development.

\begin{suppinfo}
Supporting Information is available free of charge at \url{https://pubs.acs.org/}.
\begin{itemize}
  \item Details for datasets, parameters for training target mappings, and additional figures.
\end{itemize}
\end{suppinfo}

\section*{Acknowledgements}
Funding from the Polish Science Foundation (START), the National Science Center in Poland (Sonata 2021/43/D/ST4/00920, ``Statistical Learning of Slow Collective Variables from Atomistic Simulations''), and the Ministry of Science and Higher Education in Poland are acknowledged. We thank D.E. Shaw Research for sharing Anton molecular dynamics trajectories.

\bibliography{main}

\end{document}